\documentclass[3p,times,authoryear]{elsarticle}
\usepackage{ecrc}

\volume{00}

\firstpage{1}

\runauth{Y.B. Liu et al.}

\usepackage{amssymb}

\usepackage{array}
\usepackage{graphicx}
\usepackage{multirow}
\usepackage{arydshln}
\usepackage{threeparttable}
\usepackage{booktabs}

\begin{document}

\begin{frontmatter}



\dochead{}

\title{Low-cost High Performance Distributed Data Storage for Multi-Channel Observations }


\author[a,b,c]{Ying-bo LIU}
\author[a,b,c]{Feng WANG\corref{cor1}}
\ead{wangfeng@acm.org}
\author[a]{Hui DENG}
\author[a]{Kai-fan JI}
\author[a,b,c]{Wei DAI}
\author[a,b,c]{Shou-lin WEI}
\author[a]{Bo LIANG}
\author[a]{Xiao-li ZHANG}

\cortext[cor1]{Corresponding author}

\address[a]{Computer Technology Application Key Lab of Yunnan Province,  \\Kunming University of Science and Technology, Chenggong, Kunming, China, 650500}
\address[b]{Yunnan Observatories, Chinese Academy of Sciences, Kunming, China, 650011}
\address[c]{University of Chinese Academy of Sciences, Beijing, China,100049}

\begin{abstract}
The New Vacuum Solar Telescope (NVST) is a 1-meter solar telescope that aims to observe the fine structures in both the photosphere and the chromosphere of the Sun. The observational data acquired simultaneously from one channel for the chromosphere and two channels for the photosphere bring great challenges to the data storage of NVST. The multi-channel instruments of NVST, including scientific cameras and multi-band spectrometers, generate at least 3 terabytes data per day and require high access performance while storing massive short-exposure images. It is worth studying and implementing a storage system for NVST which would balance the data availability, access performance and the cost of development. In this paper, we build a distributed data storage system (DDSS) for NVST and then deeply evaluate the availability of real-time data storage on a distributed computing environment. The experimental results show that two factors, i.e., the number of concurrent read/write and the file size,  are critically important for improving the performance of data access on a distributed environment. Referring to these two factors, three strategies for storing FITS files are presented and implemented to ensure the access performance of the DDSS under conditions of multi-host write and read simultaneously. The real applications of the DDSS proves that the system is capable of meeting the requirements of NVST real-time high performance observational data storage. Our study on the DDSS is the first attempt for modern astronomical telescope systems to store real-time observational data on a low-cost distributed system. The research results and corresponding techniques of the DDSS provide a new option for designing real-time massive astronomical data storage system and will be a reference for future astronomical data storage.
\end{abstract}

\begin{keyword}


Astronomical Databases: Miscellaneous; Methods: Massive Solar Data Storage; Techniques: High Performance I/O
\end{keyword}

\end{frontmatter}



\section{Introduction}
\label{sec:intr}
\subsection{The New Vacuum Solar Telescope}

The New Vacuum Solar Telescope (NVST) is a 1-meter vacuum solar telescope of Fuxian Solar Observatory (FSO) which is located at the northeast side of Fuxian Lake, a world-class observational site in Yunnan province of China. The main tasks of the NVST are high resolution imaging and spectral observations in both the photosphere and the chromosphere, including measurements of the solar magnetic field \citep{Liu2011,Liu2014}.

 The imaging system is a significant part of NVST.  The main structure of NVST imaging system is a multi-channel high resolution imaging system and consists of one channel for the chromosphere and two channels for the photosphere. The wavelength band for observing the chromosphere is H$\alpha$ (6563 {\AA}). The bands for observing the photosphere are TiO (7058 {\AA}) and the G-band (4300 {\AA}). So far, there are five cameras, i.e., one Andor Neo sCMOS camera (2560 $\times$ 2160 pixel with 30 frames per second (FPS)), three PCO4000 CCD cameras (4008 $\times$ 2672 pixel with 5 FPS) and one PCO2000 camera (2048 $\times$ 2048 pixel with 14.7 FPS), are installed on NVST. All channels connect to an optical splitters so that these channels can observe and record images simultaneously. The fine structures and their evolution in the photosphere and the chromosphere can be observed at the same time.

Table~\ref{tbl:expectedperf} shows the requirements of storage bandwidth and the expected performance of the NVST storage system.

\subsection{Current Computer Data Storage Technology}

 For astronomical observations, data storage technology is one of the most important issues because all of the data must be stored securely and reliably (and eventually archived intelligently) while remaining accessible on demand.

\begin{figure}[htbp]
 \begin{center}
 \begin{tabular}{ccc}
 \includegraphics[width=0.43\linewidth]{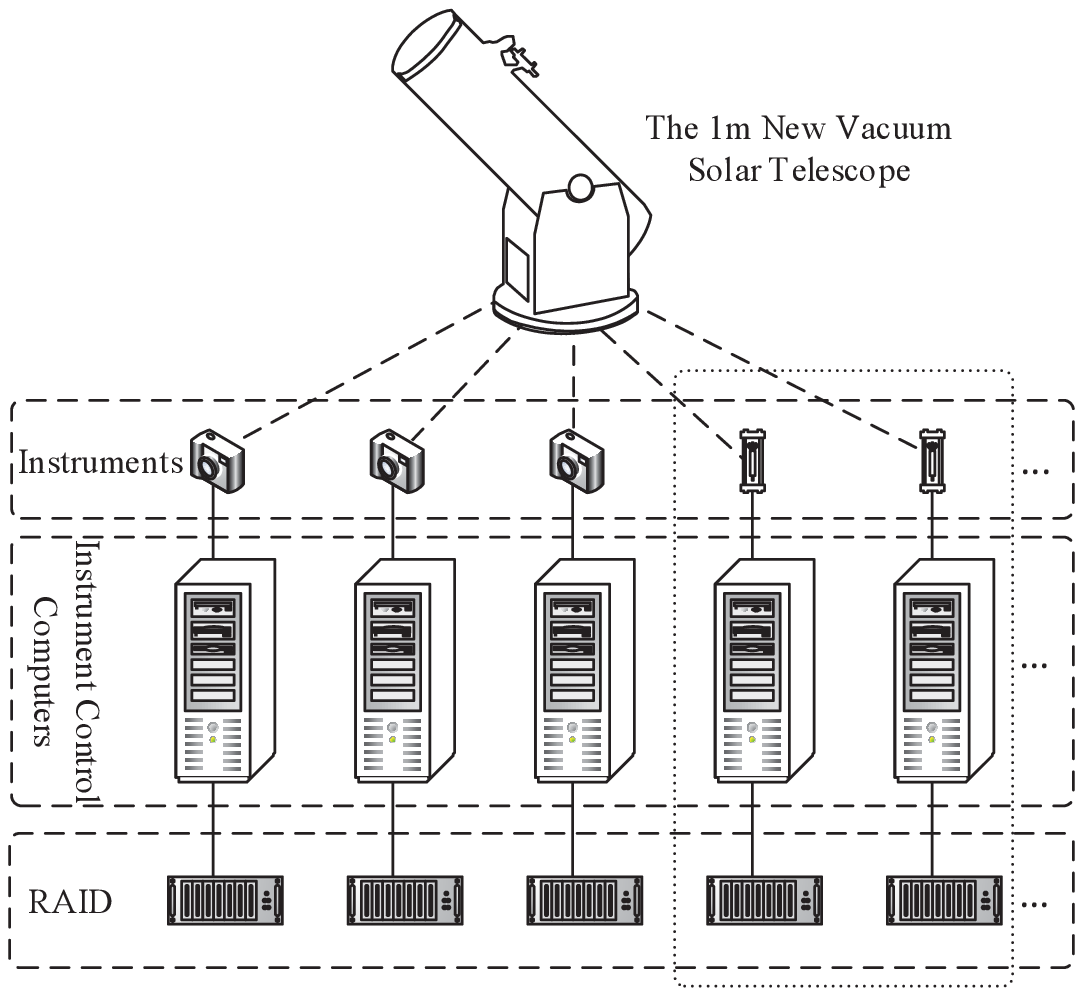} &
 \includegraphics[width=0.43\linewidth]{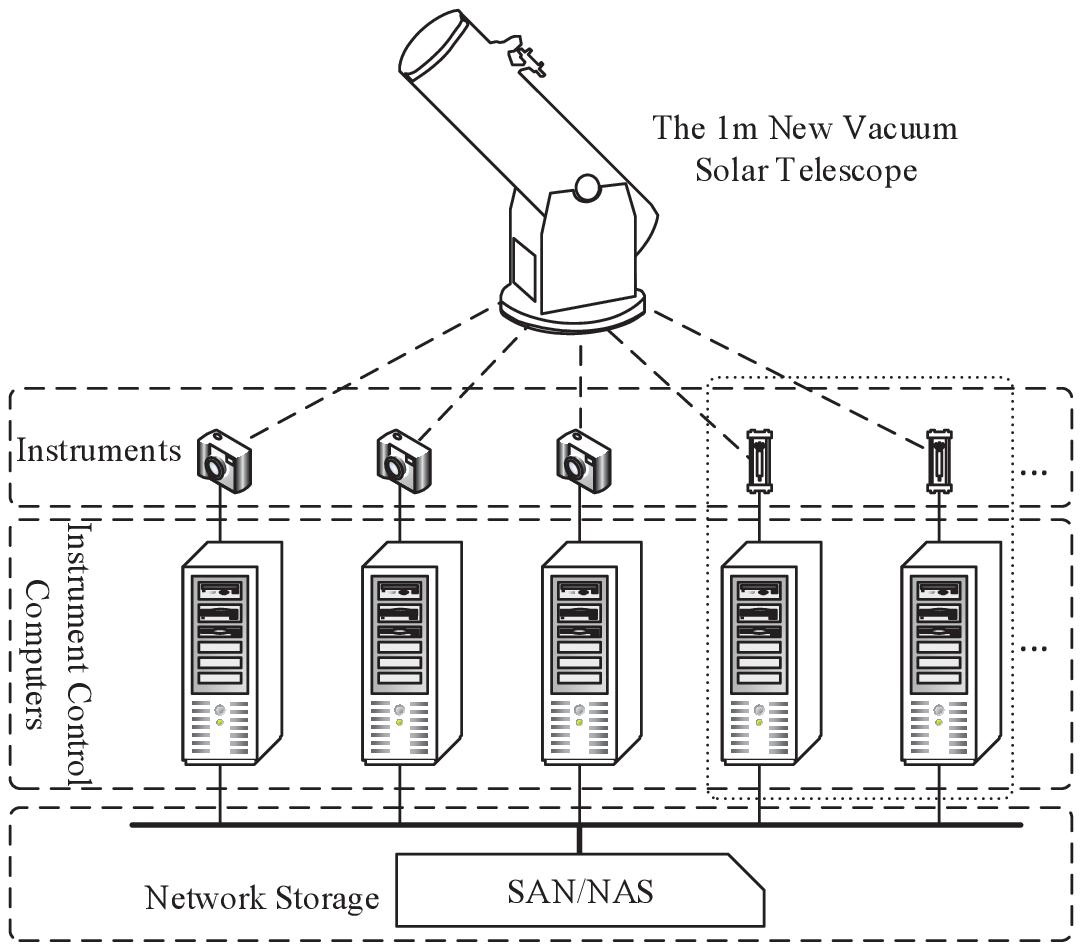}  \\
 (a)&(b)\\
 \end{tabular}
 \caption{The architecture diagram of the DSS with DAS and SAN/NAS technique. (a) shows a DAS architecture and (b) shows a common SAN/NAS architecture.\label{fig:archtecture}}
 \end{center}
\end{figure}

So far, Direct Attached Storage (DAS), Network Attached Storage (NAS) and Storage Area Network (SAN) are quite mutual. Figure~\ref{fig:archtecture} shows the architecture diagrams of the Data Storage System (DSS) with DAS and NAS/SAN technique respectively. Meanwhile, the advent of new generation storage techniques (i.e., high performance Solid State Disk (SSD) system, massive clouding storage, high speed 8 Gb and higher storage bandwidth) brings more options for the massive and high performance astronomical data storage. We investigate the previous literatures concerning the astronomical data storage and list advantages and limitations of each technique respectively in Table~\ref{tbl:comparethree}.

Besides DAS, NAS and SAN techniques, distributed storage technique is becoming a trend for massive data storage. Distributed parallel file system is such a technique to survive massive data storage; it can provide high I/O performance and extensible bandwidth, and it can be scaled out to fit for petabyte level data storage \citep{Braam2003}.
Distributed data storage systems have already been widely used in astronomy \citep{Withington2004}. Astro-WISE (http://www.astro-wise.org/) is a distributed system for astronomical data. It stores all kinds of data ranging from the raw data to the final science-ready product data \citep{Mwebaze2010}. \citet{Becla2006} utilized highly-distributed relational database and super massive databases to store astronomical data. \citet{Stonebraker2011} use non-relational database---SciDB \citep{Cudre-Mauroux2009} to store astronomical data. SciDB is already applied in radio astronomy \citep{Diepen2013}. In order to deal with the large storage requirements of MAGIC-I and MAGIC-II, \citet{Carmona2009} exploited the Distributed File System (GFS) and built a cluster of computers with concurrent access to the same shared storage units. This storage system has advantages such as reliability, flexibility and scalability. Gfarm distributed file system has been used in astronomical data analysis and visual observatory \citep{Tatebe2004}. LSST project applies Qserv system to manage large data sets, which is based on MySQL and a distributed file system---xrootd \citep{Wang2011}. However, according to the detailed investigation mentioned above, almost all distributed data storage systems currently used in astronomy mainly are focused on data archiving. Few previous studies discussed the applications of using distributed data storage technology in real-time astronomical observation.

\subsection{The Selection Of Storage Solution For NVST Data Storage System }

The data storage system is an indispensable fundamental facility of NVST imaging system. The goal of NVST storage system is that the system can satisfy the high-capacity data storage requirement now, and also be able to meet capacity needs in the future. According to the investigations and analyses in the previous subsection, it is not easy to determine which storage technique is the most appropriate for the DSS. Actually, all the storage techniques discussed have their specific advantages and limitations.

After a long-term investigation and experimental testing, we finally select distributed storage technology and build a Distributed Data Storage System (DDSS) for NVST because we believe the distributed storage technology is capable of balancing the data availability, access performance and the cost of development. The reasons for selecting distributed storage technology are listed as follows.

\begin{enumerate}[1)]
\item \emph{A Centralized Storage System.} The DDSS of NVST is a centrally administered system. The most valuable feature of the centralized storage system is that all instruments can fully share the available capacity. This is very useful for long term observation which might use different instrument or the combination of instruments. In addition, a centralized storage system would significantly reduce the maintenance workload of the observers. In general, the observers are not professionals of information technology. For them, the separated storage systems with different technical architecture are naturally difficult to maintain.

\item \emph{High Performance Storage.} The DDSS supports all instruments to acquire data with maximum acquisition speed for NVST simultaneously.  In the practical observations, it is always better to obtain faster acquisition speed for each instrument. The experimental results show the write performance of the DDSS is big enough (i.e., more than 760 MB/s while five cameras are working under full load operation) to guarantee the integral and lossless of the observational data of NVST. The DDSS supports concurrent read and write. In particular, the data read should not interfere with the performance of the data write. This is very useful for real-time data processing in observations.

\item \emph{Low-cost and Easily-extensible Capacity.} The DDSS can easily be expanded its storage capacity so as to meet the requirements of observational data storage and archive in a short period.  Meanwhile, we focus on the budget of NVST, which is always a worrisome problem in telescope construction and maintenance. The cost of the DDSS is relatively lower than that of other storage techniques.
\end{enumerate}

In this study, we focus on the design and experimental testing for implementing the low-cost and high performance DDSS for NVST. The rest of this paper is organized as follows. Section \ref{sec:ddss} briefly introduces the build of the DDSS and focuses on the experiments for testing the access performance of the DDSS. Three high performance write strategies are discussed and implemented in Section \ref{sec:application}. Discussions and a short summary are separately provided in Section \ref{sec:discussion} and \ref{sec:conclusion}.

\section{The Distributed Data Storage System for NVST}
\label{sec:ddss}
\subsection{The build of the distributed data storage system}
 After a series of preliminary experiments, we finally choose Lustre file system \citep{Braam2002} as the core system of the DDSS. Lustre is a GNU General Public licensed, open-source distributed parallel file system. The Lustre user group declared that Lustre operates best in Parallel I/O Environment and is well suited for large-scale parallel environment and capture environment \citep{Carrier2012}.

\begin{figure}[htbp]
\centering
\includegraphics[width=0.43\textwidth]{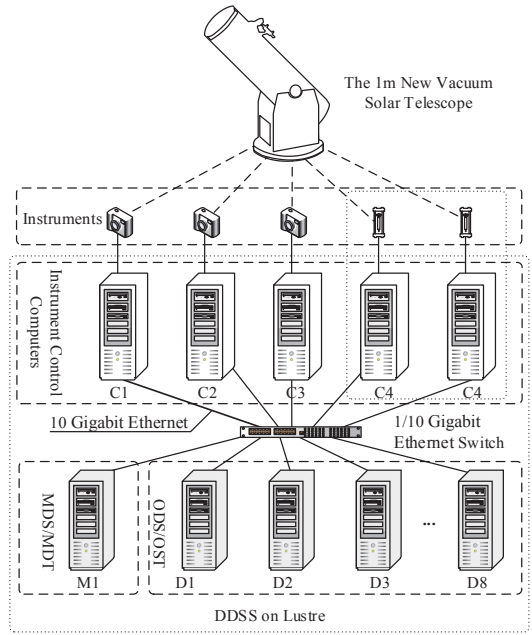}
\caption{The architecture diagram of the DDSS. Total 14 servers are integrated in the system and connected to a 1/10 Gigabit Ethernet network switch. Besides the C1 with 10 Gigabit Ethernet, all others are connected with 1 Gigabit Ethernet.\label{fig:nvstarchi}}
\end{figure}

 We build a distributed data storage system for NVST by following the installation manual of the Lustre system. Figure~\ref{fig:nvstarchi} shows an architecture diagram of the DDSS. The 1/10 Gigabit Ethernet switch has 24 $\times$ 1 Gb ports and 2 $\times$ 10 Gb ports. Five computers, which are labelled as in Figure \ref{fig:nvstarchi} are deployed as the instrument control computers. The computer (C1) having Andor sCMOS camera\footnote{http://www.andor.com/scientific-cameras/neo-and-zyla-scmos-cameras} is connected to the switch with 10 Gb bandwidth and other servers (C2-C5) are connected with 1 Gb bandwidth. 3 PCO4000 cameras\footnote{http://www.pco.de/sensitive-cameras/pco4000, unmatched fast image recording with 128 MB/s} are attached to servers (C2-C4) respectively. PCO2000 camera\footnote{http://www.pco.de/sensitive-cameras/pco2000, unmatched fast image recording with 160 MB/s} is attached to C5. These servers are deployed as client nodes of Lustre file system. Metadata Server (MDS) and Metadata Targert (MDT) are deployed in one server (M1). The other eight servers (D1-D8) are deployed as data storage nodes. Object Storage Server (OSS) and Object Storage Target (OST) are also deployed together on each server (D1-D8) respectively.

All servers of the DDSS are obtained from an outdated high performance computing cluster. Each server has 8 AMD Opteron 1.05 GHz CPU, 4 GB of memory and a 1 TB hard disk (Seagate ST1000DM003\footnote{http://www.seagate.com/www-content/product-content/desktop-hdd-fam/en-us/docs/desktop-hdd-ds1770-4-1405us.pdf}, 7200 RPM, 64 MB cache, average data rate is 156 MB/s). All machines are running CentOS 6.4 and using LVM to manage logical volume. A logical volume of 400 GB is created on the MDS used for metadata storage. MDS/MDT server has 32 CPU cores of AMD Opteron 2.40 GHz and 16 GB memory. 3 MBF2300RC \footnote{http://storage.toshiba.eu/export/sites/toshiba-sdd/media/products/datasheets/mbf2xxxrc\_datasheet\_v2.pdf} drives are configured as RAID5 by using LSI MegaRAID SAS 9260-8i controller.

Obviously, the maximum access performance of the DDSS is limited by three reasons. The first is the network bandwidth (i.e., 125 MB/s). The second is data rate (i.e., 156 MB/s) of the hard disk in each storage node. The last is the number of the storage nodes and the number of MDS/MDT servers. In the ideal situation, the maximum access performance of the DDSS with 8 storage nodes could be considered as 1,000 MB/s (i.e., 125 MB/s $\times$ 8 nodes) if we assume there are no performance overhead on MDS/MDT server.

\subsection{Performance Test}
 Considering the real scenarios of NVST observation, we conduct six kinds of experiments to test the access performance of the DDSS under different conditions respectively. We created a 1 GB temporary memory space by using {\em tmpfs} technique \citep{Snyder1990} on each instrument control computer (C1-C5). All FITS files for testing are generated in advance and stored on the {\em tmpfs}  directory. To guarantee the accuracy of the experiments, the synchronized I/O mode of Linux is activated and the write buffer function is not used.

\begin{figure}[htbp]
 \begin{center}
 \scriptsize
 \begin{tabular}{cc}
  \includegraphics[width=0.48\linewidth]{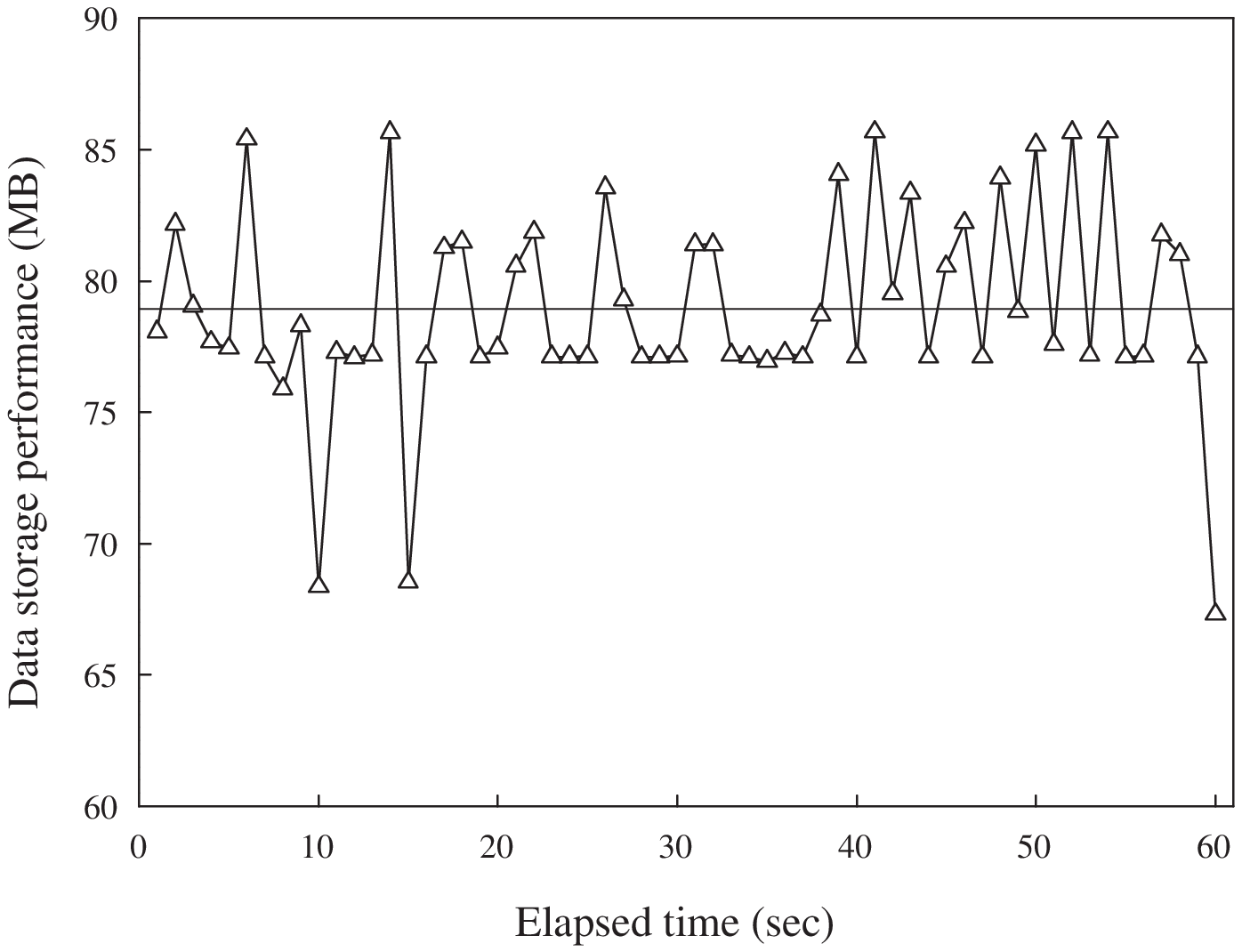} &
  \includegraphics[width=0.48\linewidth]{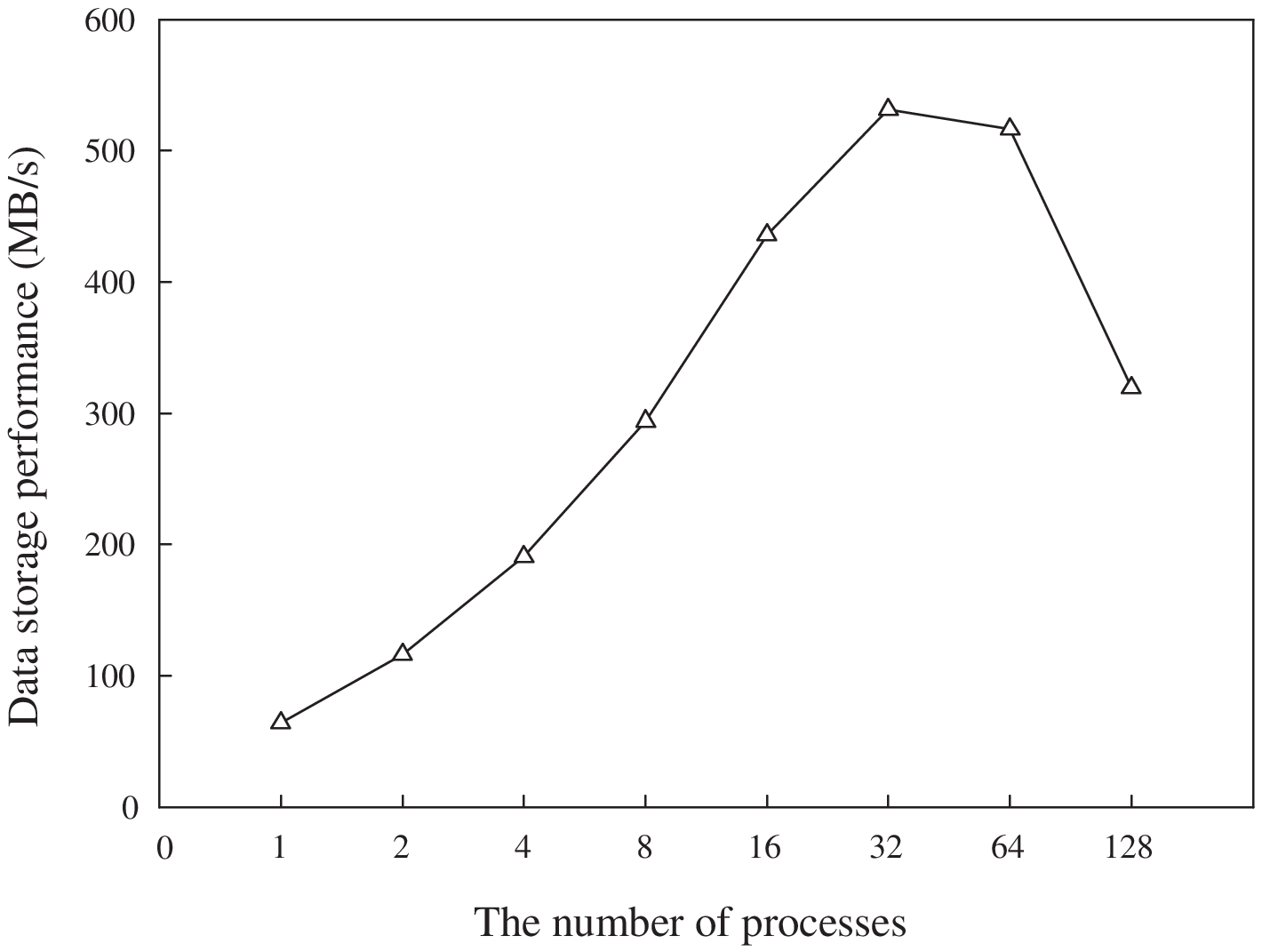} \\
 (a) & (b) \\
 \includegraphics[width=0.48\linewidth]{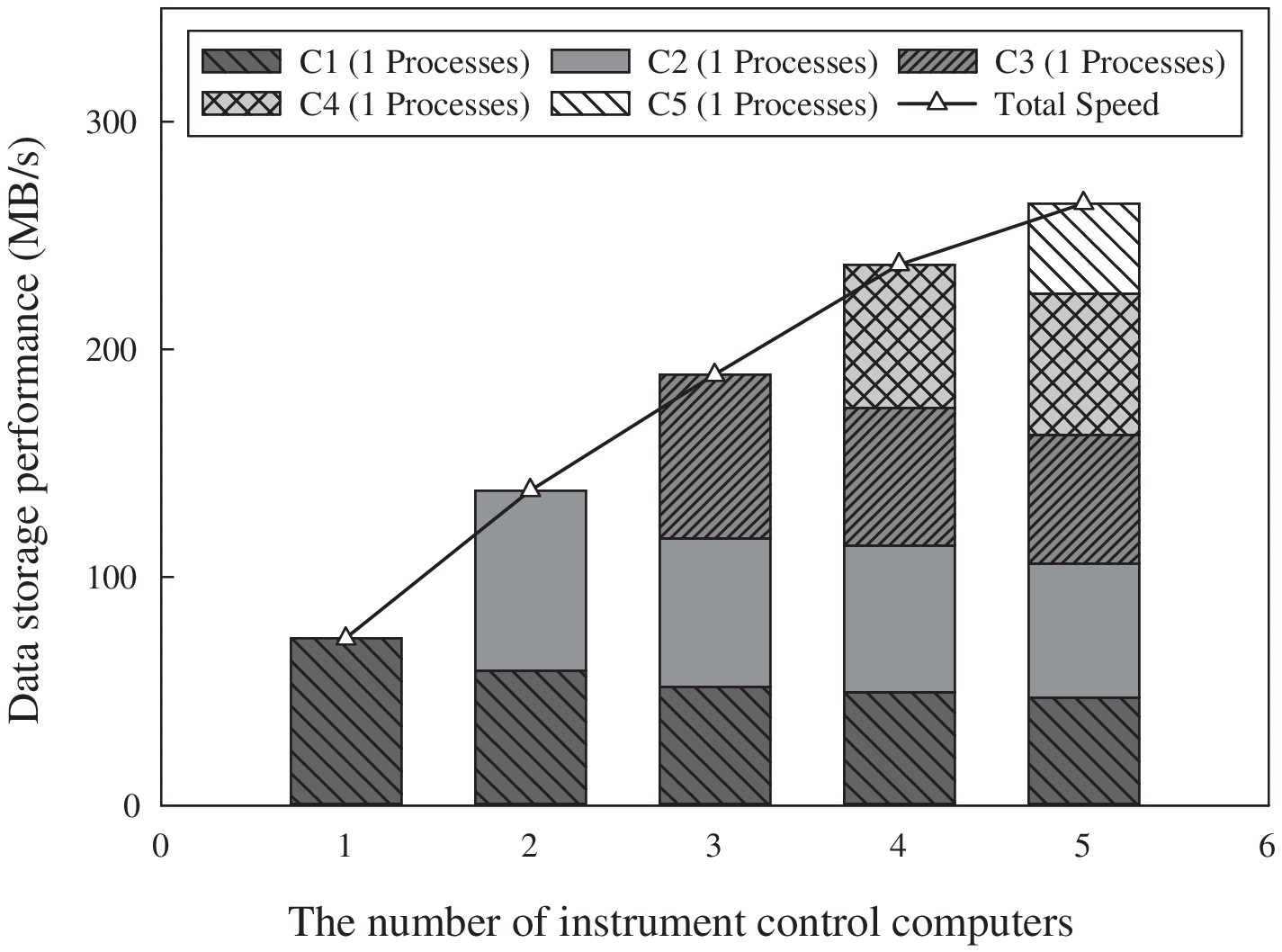} &
  \includegraphics[width=0.48\linewidth]{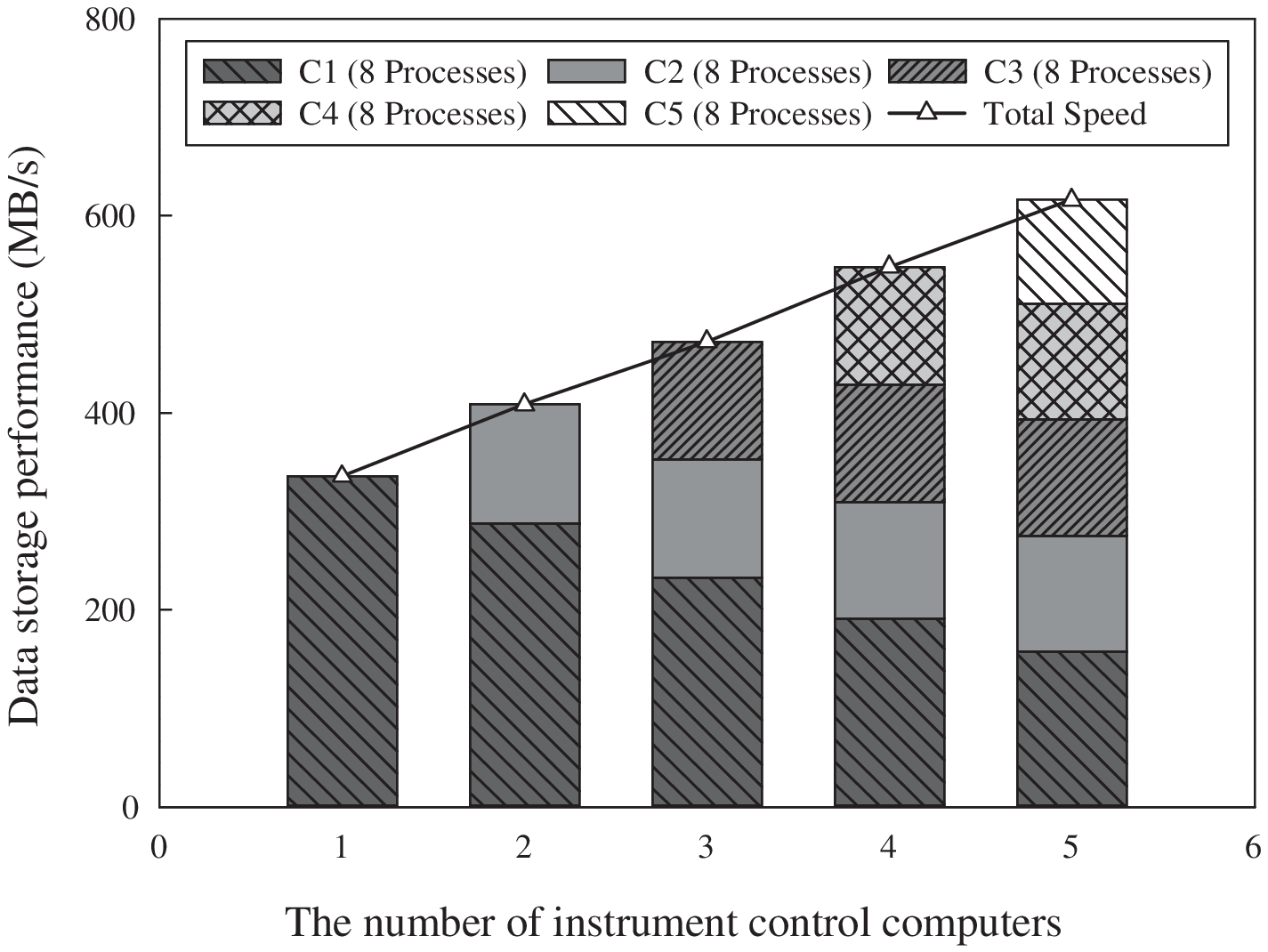} \\
 (c) & (d) \\
 \includegraphics[width=0.48\linewidth]{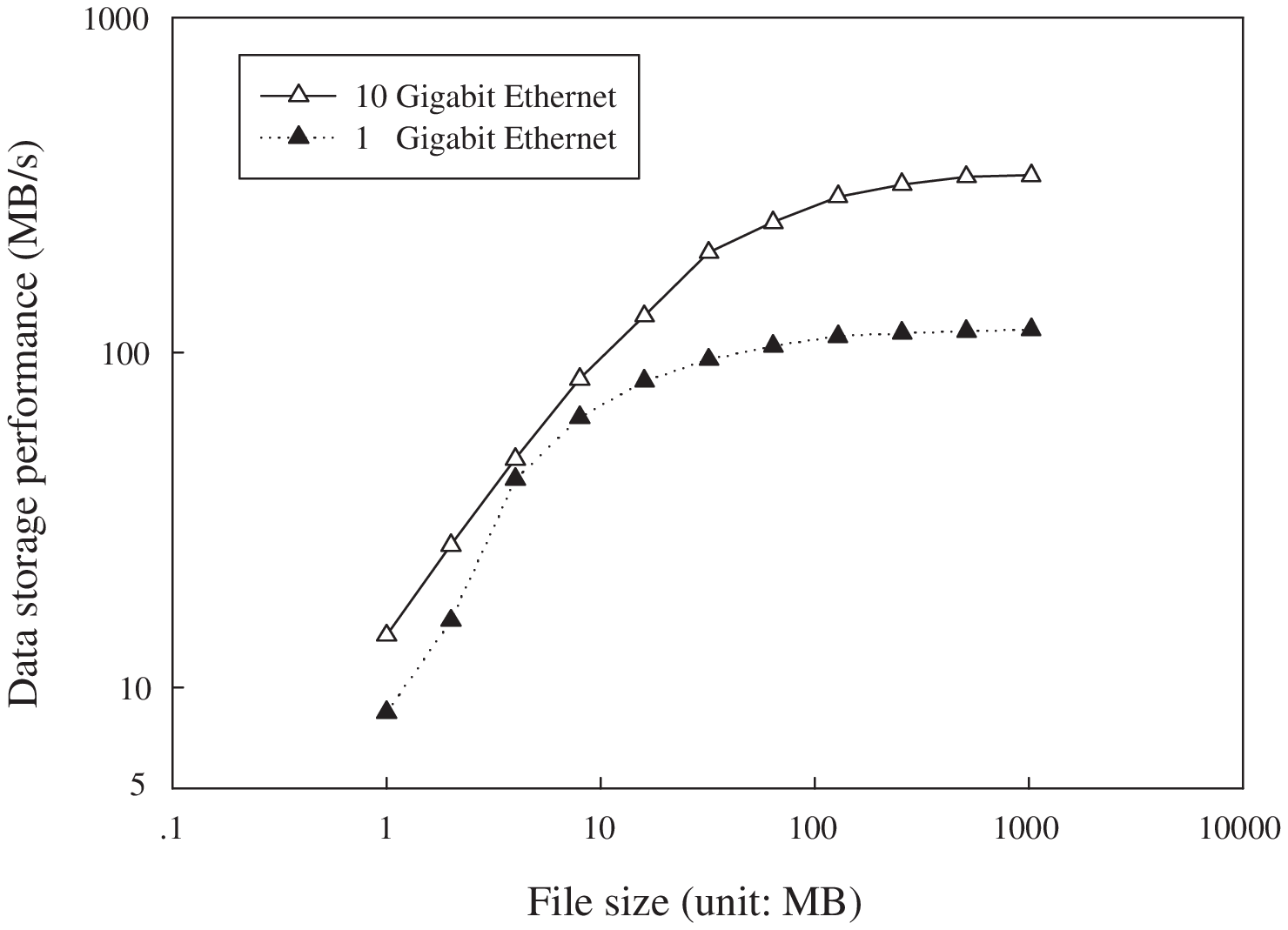} &
 \includegraphics[width=0.48\linewidth]{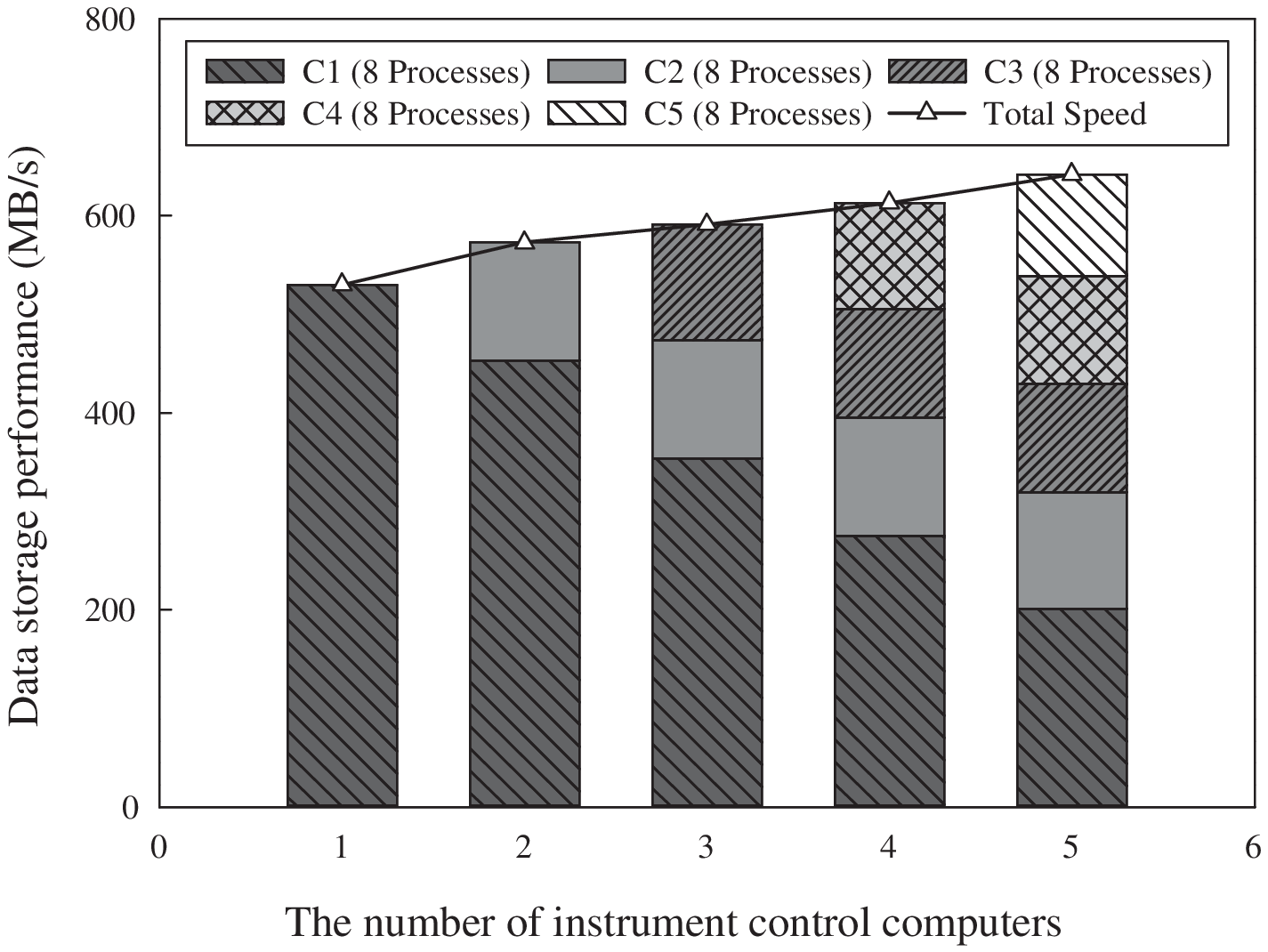} \\
 (e) & (f) \\  \end{tabular}
 \caption{The results of data storage performance testing. (a) shows the write performance of single host write; (b) shows the write performance with multi-process concurrent write on a single host; (c) shows the write performance of multi-host with single process; (d) shows the write performance of multi-host with multi-process; (e) shows the variations of storage performance with different file size by log-log plot; (f) shows the storage performance with multi-host and multi-process write with large file (256 MB).\label{fig:dssperformance} }
 \end{center}
\end{figure}

 {\em Test 1: Single Host Single-process Writing.} The single channel observation is a common mode of NVST observation. The goal of Test 1 is to investigate the maximum write performance of a single host. Considering the maximum write performance of the storage of C1 is 330 MB/s, we generated 120 FITS files with 8 MB on the {\em tmpfs} in advance. We design a test program to write the files on the DDSS repeatedly. We record the variation of network traffic for 60 seconds. Figure~\ref{fig:dssperformance} (a) shows the variations of write performance. The write speed is only approximate 80 MB/s which is far below our expectation of 330 MB/s.

 {\em Test 2: Single Host Multi-process Writing.}  As for the undesirable result of the Test 1, we further investigate the performance test literatures of Lustre to analyze the causes. We notice that all the tests on Lustre are seriously dependent on multi-process technique to write data in parallel.  Therefore, we test the write performance under the condition of single host multi-write. The testing program generates different numbers of process to simultaneously write the files to the DDSS respectively. We also record the variations of network traffic during 60 second and further calculate their average values. The results are shown in Figure~\ref{fig:dssperformance} (b). Comparing to the result of Test 1, the maximum write performance in Test 2 is greater than 500 MB/s while 32 processes running simultaneously.

 {\em Test 3: Multi-host Single-process Writing.}  The goal of Test 3 is to investigate the write performance while multiple hosts write in parallel. The experimental environment is as same as Test 1. Five computers (C1, C2, C3, C4 and C5) are used to write data in parallel. We record the network traffic at Network Interface Card (NIC) of each computer for 60 seconds and calculate the average network traffic under the conditions of C1, C1-C2, C1-C3, C1-C4 and C1-C5 respectively (see Figure~\ref{fig:dssperformance} (c)). Obviously, the total performance is only about 280 MB/s which is too poor to meet the requirements of the multi-channel observations.

 {\em Test 4: Multi-host Multi-process Parallel Writing.} The goal of Test 4 is to investigate the write performance while multiple hosts write data with multiple processes in parallel. The experimental environment and the experimental procedure are as same as Test 3. The difference is there are eight processes write simultaneously the files to the DDSS instead of only 1 process in Test 3. Figure~\ref{fig:dssperformance} (d) shows the final results of the Test 4. Obviously, although the total performance reaches more than 600 MB/s, the performance of C1 is about 320 MB/s.

 {\em Test 5: Optimal File Size.} The goal of Test 5 is to investigate the impact of file size on high performance data storage. Based on the results of Test 1 and Test 3, we realize that there is something that slows down the storage speed. According to the previous related studies \citep{Baker1991, Yu2008, Howison2012}, the file size is one of the most significant factors in data storage. Therefore, we test the write performance under the condition of different file size respectively. We use C1 and C2 to write files with the file size of 1 MB, 8 MB, 32 MB, 64 MB, 128 MB, 256 MB, 512 MB and 1024 MB respectively. Two results are shown in Figure~\ref{fig:dssperformance} (e). Obviously, the increasing of the file size is an effective approach to decrease the frequency of the file open and file close. For both of 10 Gb Ethernet and 1Gb Ethernet, the results show that the write speed reached the theoretical maximum value when the file size is larger than 256 MB.

{\em Test 6: Multi-host Multi-process Parallel Writing With Large File.} The Test 6 is similar to the Test 4. The difference between the Test 4 and Test 6 is the file size. In Test 6, the size of each file is 256 MB. Figure~\ref{fig:dssperformance} (f) shows the final results of the Test 6. The total access performance reaches more than 600 MB/s which is the highest among all tests. Meanwhile, the write performance of C1 reaches more than 400 MB/s which is greater than the acquisition requirement of 330 MB/s.

\subsection{Experimental Results Analysis}

 Six tests present a series of valuable quantitative analysis results and implications for designing the DDSS.

\begin{enumerate}[1)]
 \item The distributed storage system has ample capacity to support single and multiple hosts to store data simultaneously. The total performance of 600 MB/s in Test 4 and Test 6 definitively proves that the distributed storage system can provide excellent performance for real-time data storage. However, using conventional single process programming cannot obtain high performance of data write.

 \item The experimental results show the DDSS is able to provide basically reliable performance of data write regardless of any conditions such as single host single-process write, single host multi-process write, multiple hosts single-process write and multi-host multi-process write. This is more important for continuously data write which is demanded urgently in astronomical observation.

 \item The file size is an important factor in distributed data storage. The write performance of small size file is quite poor. However,  the observational data of NVST are stored as FITS format. For the aggregation of multi FITS files, it should be fully considered the definitions.

 \item To support multi-channel simultaneous observations, it is necessary to use multi-process or multi-thread technique to increase the write performance. It means the previous single-process instrument control software has to be redesigned and rewritten.

 \item The write performance of  C1 would decrease significantly under multi-host and multi-process (see Figure~\ref{fig:dssperformance} (d) and Figure~\ref{fig:dssperformance} (f)). However, the write performance variations of other computers are not obvious. It is worth studying further.
\end{enumerate}

\section{The Application of the DDSS in Real Observations}
\label{sec:application}

 According to the experimental results in the previous section, we rewrite the instrument control software and deploy on each instrument control computer (C1-C5). The software is written by C++ language and runs on the Linux platform. Two storage strategies, i.e., multi-process simultaneous write and large size file, are implemented and integrated into the instrument control software.

\subsection{Multi-Process Simultaneous Writing Strategy}
 Multi-process concurrent write would fully utilize the available I/O bandwidth so that it can effectively improve the write performance in data storage. The multi-process write simultaneously is simply programmed in real software as follows.

 \begin{enumerate}[1)]
 \item  The main process of the software is in charge of data acquisition. It would receive the data captured from the camera, convert to the FITS format and save to a buffer on computer memory.
 \item  After at least 8 files are generated on the memory, the other 4 or 8 processes begin to write each buffer to the DDSS simultaneously. Meanwhile, the main process is still working so as to keep the data acquisition.
\end{enumerate}

 In real observation with multi-process simultaneous write strategy, we analyze the results on single channel observation (C1), multi-channel simultaneous observation (C1, C2, C3, C4 and C5). Figure~\ref{fig:parallelperformance-a} shows a histogram of storage performance, and the real FPSs under different observational modes are shown in Table~\ref{tbl:parallelperformance-b}. It is no doubt that the multi-process simultaneous write is an effective way to increase the storage performance. However, the limitation is also obvious. Even if only one instrument (e.g., C1)  is in observation, the FPS cannot reach its maximum specifications. The maximum FPS of Andor Neo sCMOS camera is 30 FPS. But the actual value in observation is only 26.9 FPS.

\subsection{Large File Aggregation Strategy}

 According to the experimental results of Section 2, the large file (256 MB or more) is a necessary precondition for high performance storage. Therefore, it is necessary to aggregate many small FITS files into a big one. Actually, the aggregation of multi FITS files is not an easy job. If we simply append a FITS file to another, it is easy to combine multiple files into a big one. However, this would lead to disastrous consequences.  Astronomical data process software cannot recognize and open the FITS file directly. And we have to split original FITS files one by one from the aggregated file while conducting post processing of these FITS files.

\begin{figure}[!htbp]
 \begin{center}
  \scriptsize
 \begin{tabular}{c:c}
 \includegraphics[width=0.46\linewidth]{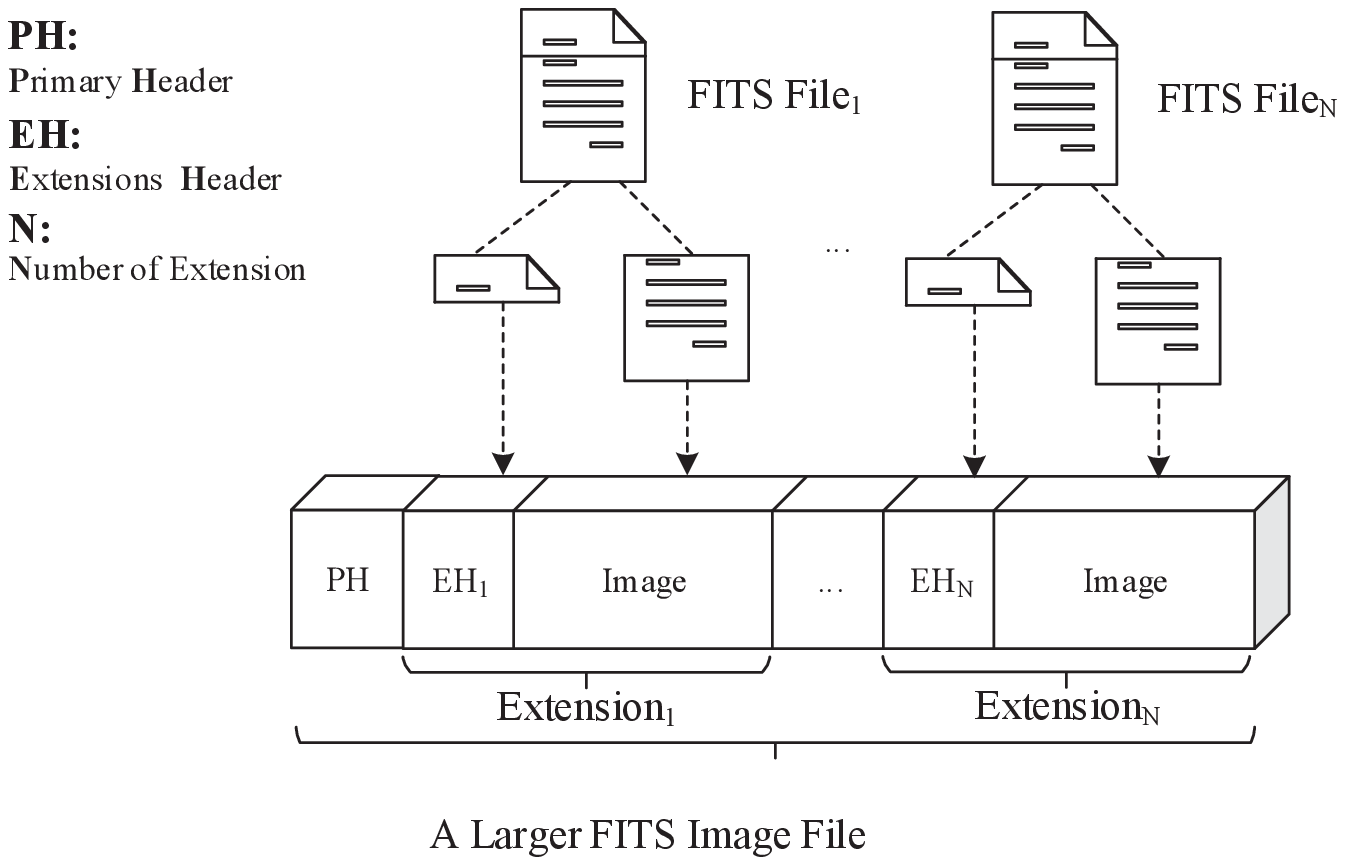} &
 \includegraphics[width=0.46\linewidth]{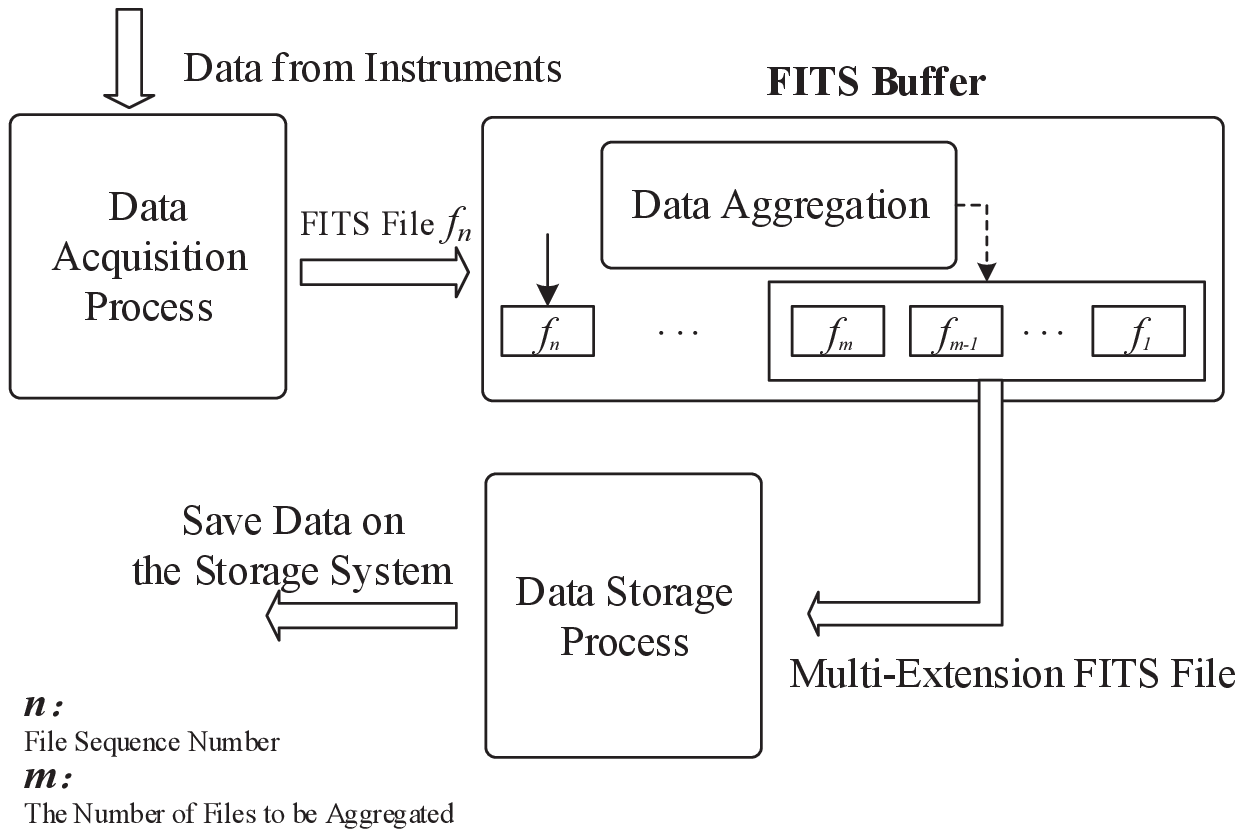} \\
 (a) & (b)
 \end{tabular}
 \caption{ (a) The schematic diagram of FITS file aggression. $N$ means the number of the files needed to be aggregated which should be different in different instrument. For example, for Andor Neo camera, $N$ should be 256/8 = 32. (b) The data flow diagram of large file aggregation strategy.\label{fig:assemblyfits}}
 \end{center}
\end{figure}

 To aggregate multi FITS files in real observation, we implement an Automatic FITS File Aggregation (AFA)  function in the instrument control software. Figure \ref{fig:assemblyfits} shows a principle diagram of the AFA that fully utilizes the definition of extension format in FITS standard \citep{Pence2010,Smith2011}. The AFA is running on a separate process to consistently monitor the files on a memory file system which is generated by {\em tmpfs} technique. When it is detected that adequate files have been generated and the amount of the file size is larger than a specified size (e.g., 256 MB), the AFA would aggregate these files and generate a single FITS file.
	
 Figure~\ref{fig:afaperformance-a} shows the real performances while C1 or multi-host acquiring images simultaneously.  The FPS of C1 can reach 28.5 (see Table~\ref{tbl:afaperformance-b}), and the value is close to the maximum FPS of Andor Neo sCMOS camera. We can see that single channel of C1 has better performance than that in Figure~\ref{fig:parallelperformance-a}. But with more channels (C2-C5) in use, the total FPS of image acquisition is lower than that in Figure~\ref{fig:parallelperformance-a}.

\subsection{Hybrid Strategy High Performance Writing}

 The results of previous two subsections bring us more idea for improving access performance. Two strategies, such as multi-process and large file, might improve the write performance. However, their mechanisms for improving performance are different. Multi-process aims to fully utilize the communication bandwidth. The large file would significantly decrease the frequency of file open and close. Therefore, the combination of two strategies enable us to obtain further improvement in the I/O performance.

 We implement a hybrid storage strategy which integrates multi-process strategy and large file strategy. To meet the actual requirements of real observations, a computer with 1 Gigabit Ethernet network is used to read the observational data in parallel while other instruments are in observations. Figure~\ref{fig:hbperformance-a} shows the optimal results of the DDSS after many experiments (Figure~\ref{fig:dssperformance}, \ref{fig:parallelperformance-a} and \ref{fig:afaperformance-a}). In hybrid strategy, the FITS files are aggregated into a file size of 64 MB and there are other 5 processes running on the background to store the aggregated file.

 Comparing to the previous two strategies, the hybrid strategy manifests more advantages. On the one hand, the real FPS of the instrument reaches the maximum FPS which is declared by the manufacture (see Table~\ref{tbl:hbperformance-b}). On the other hand, the variation of write performance under the hybrid mode would not drop significantly and is more stable while in multi-host simultaneous observations. The most important, the simultaneous data read has no significant impact on the write performance.

\section{Discussion}
\label{sec:discussion}
 According to the investigation of previous literatures, few astronomical telescopes store their real-time observational data into a distributed storage system. The reason is no one can believe the hardware reliability and the I/O performance of the distributed storage system. Hence, although the distributed data storage system has been widely used in the field of astronomy, the main application of the distributed storage system is for data archiving and post-processing.

In our study, we attempt to build the distributed storage system into NVST data process system for real-time observation and data post-processing. Once the DDSS is running, the system shows the distinguish advantages beyond expectations. After the real application of the DDSS, there are several issues and limitations needed to be discussed.

\begin{enumerate}[1)]
\item {\em Scalability.}  Figure~\ref{fig:hbperformance-a} shows that total write and read performance is 630.7 MB/s, and the total pure write performance of 5 channels is 580.2 MB/s, which is only 58\% of our expected performance of 5 channels write in parallel. It also shows that the average FPS of 5 channels can achieve at 58\% of the maximum of our required image rate. We believe better performance can be obtained by carefully tuning the Lustre system. Obviously, the DDSS with total 8 servers has proved that the performance is suitable for real-time data storage of modern astronomical telescopes. We can expect that the DDSS with more storage nodes can also meet the requirements of using more channels and cameras in the near future. For example, if 8 new PCO4000 cameras with average FPS of 5 are installed, the total transport speed rate would be $8 \times 105 = 840$ MB/s. The performance can be supplied by adding extra 12 data nodes (840MB / 58\%  / 125 MB = 11.6). Thus, in theory, the DDSS that can support 13 observational channels working in parallel should have at least 20 data nodes.

 The MDS is an important part in the DDSS. The amount of memory used by the MDS determines the number of clients connected to the system. Referring to the Lustre manual (http://wiki.lustre.org/images/3/35/821-2076-10.pdf), 4 GB RAM can serve a single MDT on an MDS with 1,000 clients, 16 interactive nodes. There,  the hardware configuration of MDS in our study is enough for 13 channels. Similarly, the OSS memory of our system can also provide the required performance.

\item {\em Manageability And Usability.} In general, the application and deployment of open source software seriously depend on the professionals because of limited documents and supports. Although after several years software development, the current Lustre system has been proved a mutual system. However, the deployment of Lustre is still a challenging work. The administrators of the DDSS must understand the fundamental concept of distributed computing, metadata server, metadata target, object storage server and object storage target. There is no free universal management software for administrators to monitor the full system. The administrators have to remember the commands and maintain the DDSS via command line interface under Linux terminal. But overall, although there is still a gap between open source Lustre system and the other commercial products, the manageability of Lustre is acceptable by the astronomers of NVST after a short-term training.

\item {\em High Performance Storage.} Although Lustre claims that it can provide end-to-end throughput over the Gigabit Ethernet network in excess of 100 MB/s and 1,000 MB/s over 10 Gigabit Ethernet network, these remarkable results are always obtained through multi-process write on a specific test environment. According to our experiments, it is not easy to reach such storage speed because the astronomical observation is a continuous process of alternating data acquisition and data storage. Many factors play important roles in the high performance storage of the Lustre.

\item {\em Fault Tolerance and Reliability}. The reliability of the DDSS is considered as error-free and maintenance free running in the observation. Comparing to the traditional storage techniques, reliability is the biggest concern for building the DDSS of NVST. To test the reliability of the DDSS, we simulate the real observations and continuously write the data to the DDSS during total 168 hours ($7~days~\times~24~hours$) without any hardware and software maintenances. The system successfully and easily passed the testing.

The Metadata Server and Metadata Target are the core components of Lustre distributed file system. In the current DDSS, we only setup one MDS/MDT server to store metadata. It means that the DDSS has a problem of a single point of failure and is not fault tolerance. However, the improvement of the reliability of the DDSS would cost too much because we have to add power management software, hardware and high availability (HA) software.

\item {\em Low-cost and Easily-extensible Capacity.} The most surprising finding in using Lustre in astronomical observation is the low-cost. For many high performance storage system, the costly 8 Gb HBA and high speed SAS disks are the indispensable parts. However, the DDSS only consists of low-cost SATA hard disks and a Gigabit Ethernet network. It means that with the support of Lustre system, the system composed by these low-cost equipments reaches the high performance which commonly obtained by the expensive devices.
\end{enumerate}

It should be pointed out, the distributed data storage system is not a new technique to displace the traditional storage techniques. On the contrary, the distributed data storage technique is the further applications and integrations of traditional storage techniques. For example, the MDT/MDS server can use the SAN technique to enhance its write performance and the reliability. The OSS/OST can use RAID technique to further improve the access performance. Actually, the combination of these storage techniques would bring us more outstanding features such as higher performance and higher availability.

\section{Conclusion}
\label{sec:conclusion}
 The storage of massive observational data is a critically significant issue for modern telescopes. In this study, after requirements analyses for the storage system of NVST, we build a distributed storage system for real-time observational data storage based on the Lustre distributed file system. The system has been deployed in FSO and has been carefully tested under real-time astronomical observations. Overall, our study has the following contributions and can be referenced by other telescopes in their data storage issues.

\begin{enumerate}[1)]
 \item This is the first attempt to use a distributed storage system in real-time astronomical observation for a modern solar telescope. Practices show that the low-cost distributed storage system can support multi-host simultaneous write with high performance and meet the requirements of NVST real-time observations. This would be referenced by other modern telescopes, especially the telescopes with multi-channel simultaneous observations.

  \item Series of experiments has been conducted and the experimental results show that the number of simultaneous write and the file size are two key factors for improving the performance of data storage. Meanwhile, the different storage strategies have significant impacts on the access performance of the distributed data storage system.

  \item The storage of the observational data is a complex engineering problem. The storage system and the instrument control system are both significant for designing high performance data storage system. Our study proves that the storage strategy has significant impact on access performance of the storage system. A well-designed and robust instrument control software is the premise of the high performance storage system.
\end{enumerate}

 Overall, the distributed data storage technique is indeed a perfect solution for massive and high performance data storage. The research results and corresponding techniques of the DDSS might provide a new option for designing real-time massive astronomical data storage system and will be a reference for future astronomical data storage.

\section*{Acknowledgements}
 This work is partially supported by the National Natural Science Foundation of China (No. U1231205, 11263004, 11203011, 11163004 and 11103005) and Natural Science Foundation of Yunnan Province (No. 2013FA013, 2013FA032, 2013FZ018). The authors also gratefully acknowledge the helpful comments and suggestions of the reviewers.

\bibliography{bib}
\bibliographystyle{elsarticle-harv}


 \begin{table}[htbp]
\caption{Cameras bandwidth requirements and the expected performance of the storage system\label{tbl:expectedperf}}
\scriptsize
\tabcolsep 3pt
\renewcommand{\arraystretch}{1.3}
\vspace*{2mm}
\centering
\begin{threeparttable}
\begin{tabular}{p{35pt}p{40pt}p{40pt}p{35pt}p{45pt}p{40pt}p{35pt}}
\hline
Camera& Max. network bandwidth $V_b$ (MB) &Average HDD data rate $V_d$ (MB/s)&Size per frame $S_f$ (MB) & Expected transport rate $V_t$ (MB/s)&Expected image rate $V_i$ (FPS)\\
\hline
Andor Neo&1,250 & 156 & 11  & 330 & 30 \\
PCO4000  & $125 \times 3 $   & $156 \times 3 $ & 21 & $105 \times 3 $   & $5 \times 3 $ \\
PCO2000  &125  & 156 & 8  & 117.6 & 14.7\\
Total & & & & 762.6 &\\
\hline
\end{tabular}
\begin{tablenotes}
    \item [~] Notes: each pixel is stored using two bytes; $V_t = S_f \times V_i$.
\end{tablenotes}
\end{threeparttable}
\end{table}

\begin{table}[htbp]
\begin{center}
\caption{Advantages and limitations of three storage techniques\label{tbl:comparethree}}
\small
\tabcolsep 3pt
\renewcommand{\arraystretch}{1.3}
\vspace*{2mm}
\begin{tabular}{c|p{180pt}|p{180pt}}
\hline
Techniques & \centerline{Advantages} & \centerline{Limitations} \\
\hline
DAS  &  1) Simple, low cost and easy to setup, manage and supervise \citep{Sacks2001}. Ideal for localized files sharing in environments with a single server. 2) Dedicated storage bandwidth and easily setup a high performance system. & 1) Management complexity with the addition of new servers, separately storage management for each server. 2) Unshared available storage capacities with each other for instruments (see Figure~\ref{fig:archtecture}). 3) Limited storage space and hard to extend. \\
\hline
NAS & 1) Optimized for ease-of-management and file sharing using lower-cost Ethernet-based networks. Relatively quick installation. 2) Automatically storage capacity assignment on demand. Shared storage across multiple servers \citep{Sacks2001}. Widely used in astronomical data archiving \citep{Barbieri2003,Balard2006}. & 1) Inefficient storage speed and hardly meet the requirement of NVST high performance storage. 2) Sharable bandwidth and not suitable for data transfer intensive applications (see Figure~\ref{fig:archtecture}).  3) Complex backup, need to take backup of each server. \\
\hline
SAN & 1) Optimized for performance and scalability. High performance storage with 8 Gb or higher fibre channel media \citep{Sacks2001}. 2) Widely used in astronomy community (e.g., \citet{Li2007,Keller2003,Norton2008} )  &  1) High degree of sophistication, expensive \citep{Gray2000}. 2) Require specialized knowledge and training to configure and maintain. Need to manage user privileges, file locking and other security measures. \\
\hline
\end{tabular}
\end{center}
\end{table}

\makeatletter\def\@captype{figure}\makeatother
\begin{minipage}{0.5\textwidth}
\centering
\includegraphics[scale=0.56]{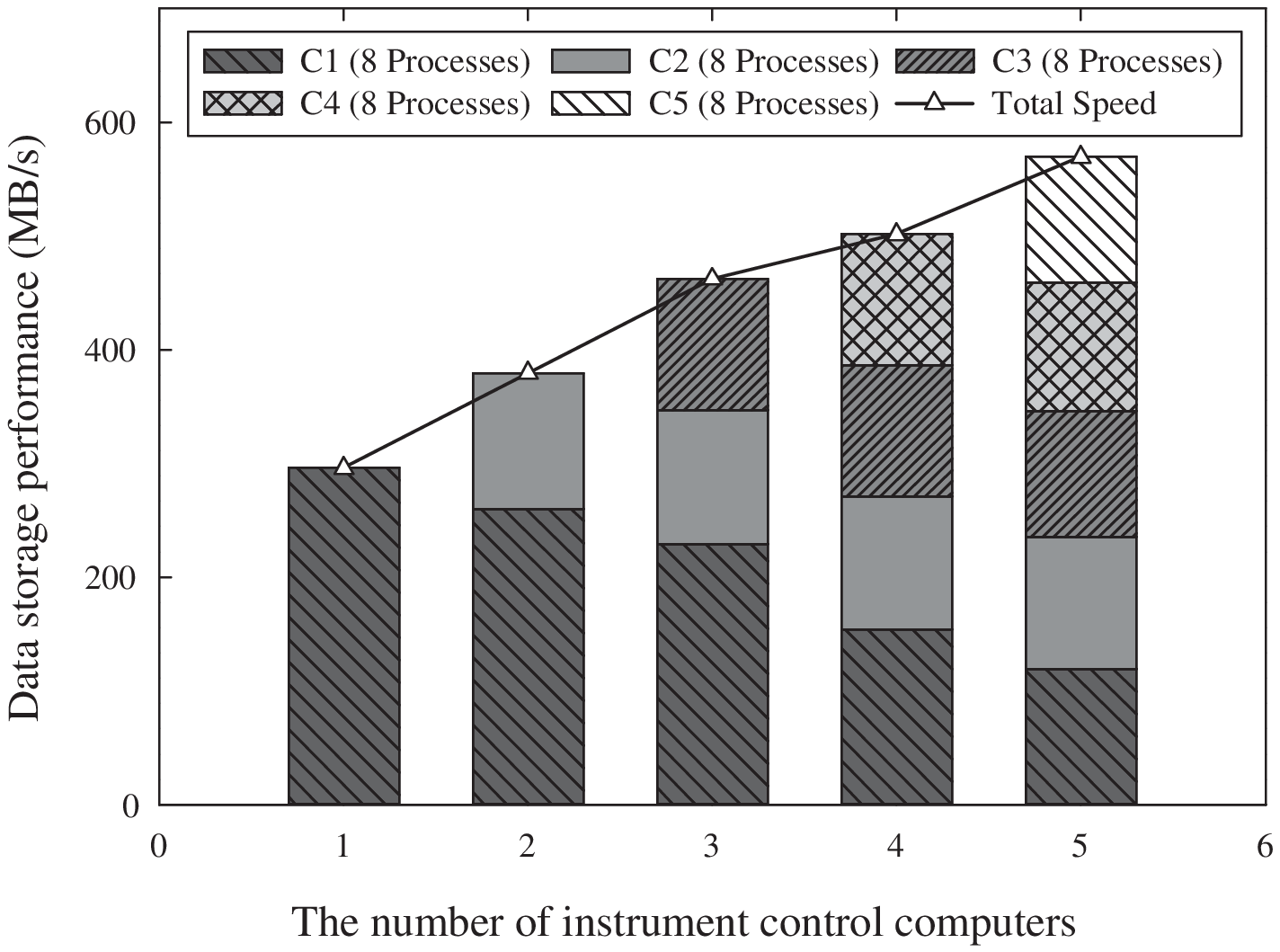}
 \caption{The performance of the storage with multi-process simultaneous write strategy in real observation.\label{fig:parallelperformance-a}}
\end{minipage}
\makeatletter\def\@captype{table}\makeatother
\begin{minipage}{0.45\textwidth}
 \caption{Instruments data acquisition performance (unit: FPS) and the ratio of FPS (Real FPS / Maximum FPS)\label{tbl:parallelperformance-b}}
\centering
\scriptsize
\tabcolsep 1pt
\renewcommand{\arraystretch}{1.3}
\vspace*{2mm}
\begin{tabular}{>{\centering}m{0.9cm}>{\centering}p{1.14cm}>{\centering}p{1.14cm}>{\centering}p{1.14cm}>{\centering}p{1.14cm}>{\centering}p{1.14cm}p{0.75cm}<{\centering}}
  \hline
  Cameras & Andor Neo (FPS) & PCO4000 (FPS) & PCO4000 (FPS) & PCO4000 (FPS) & PCO2000 (FPS)  & Total (FPS) \\
  \hline
  \multirow{4}{*}{Rates} & 26.9 (90\%) & --- & --- & --- & --- & 26.9 \\
   & 23.6 (79\%)  & 5.0 (100\%) & --- & --- & --- & 28.6 \\
   & 20.8 (69\%)  & 5.0 (100\%)  & 5.0 (100\%) & --- & --- & 30.8 \\
   & 14.0 (47\%)  & 5.0 (100\%)   & 5.0 (100\%)  & 5.0 (100\%) &---& 29.0 \\
   & 10.9 (36\%)  & 5.0 (100\%)   & 5.0 (100\%)  & 5.0 (100\%) & 13.8 (95\%) &39.7 \\
  \hline
\end{tabular}
\end{minipage}

\makeatletter\def\@captype{figure}\makeatother
\begin{minipage}{0.5\textwidth}
\centering
\includegraphics[scale=0.56]{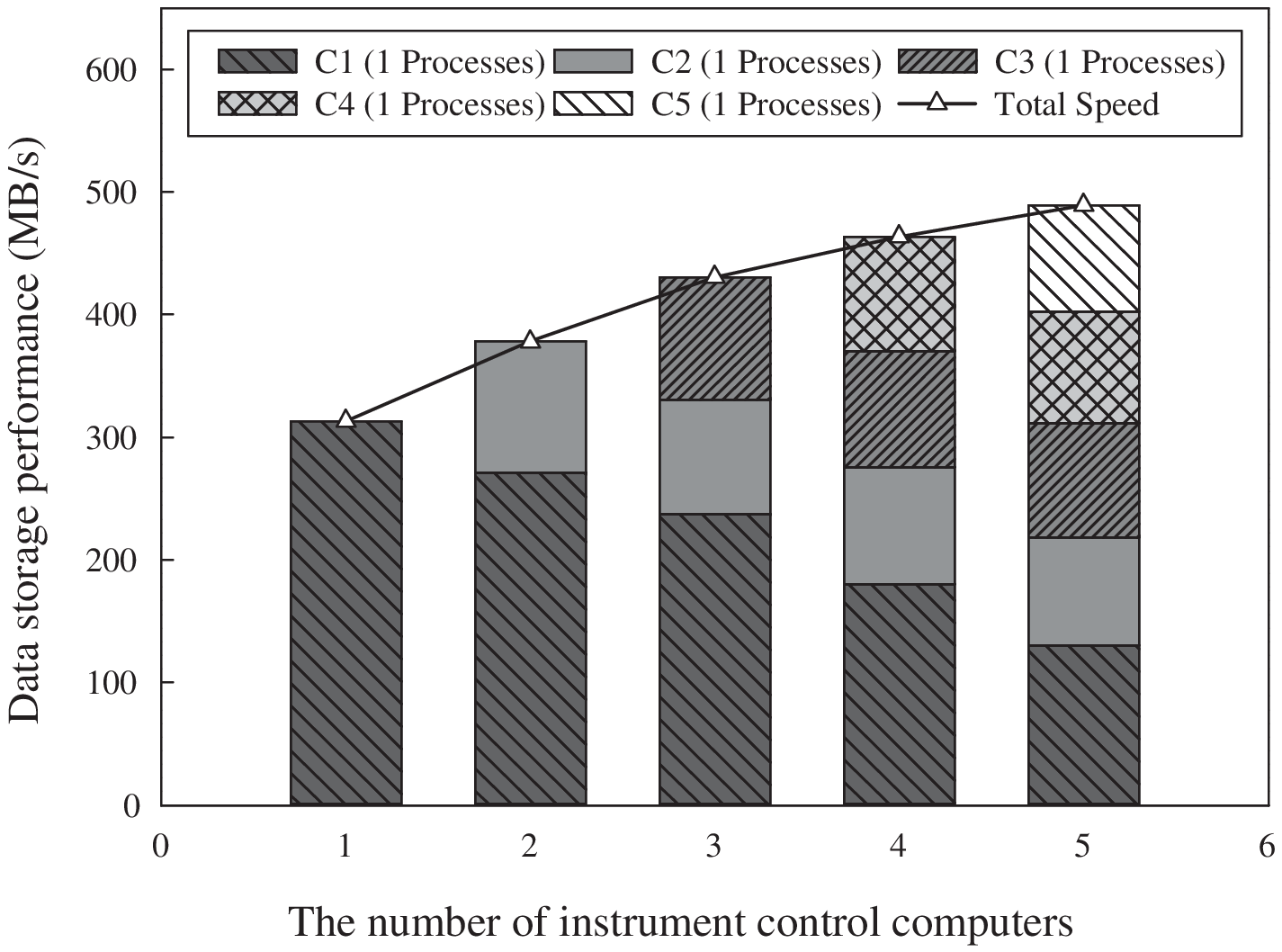}
 \caption{Real parallel write performance using aggregated FITS files write strategy in real observation.\label{fig:afaperformance-a}}
\end{minipage}
\makeatletter\def\@captype{table}\makeatother
\begin{minipage}{0.45\textwidth}
\caption{Instruments data acquisition performance (unit: FPS) and the ratio of FPS (Real FPS / Maximum FPS)\label{tbl:afaperformance-b}}
\centering
\scriptsize
\tabcolsep 1pt
\renewcommand{\arraystretch}{1.3}
\vspace*{2mm}
\begin{tabular}{>{\centering}m{0.9cm}>{\centering}p{1.14cm}>{\centering}p{1.14cm}>{\centering}p{1.14cm}>{\centering}p{1.14cm}>{\centering}p{1.14cm}p{0.75cm}<{\centering}}
  \hline
  Cameras & Andor Neo (FPS) & PCO4000 (FPS) & PCO4000 (FPS) & PCO4000 (FPS) & PCO2000 (FPS)  & Total (FPS) \\
  \hline
  \multirow{4}{*}{Rates} & 28.5 (95\%) & --- & --- & --- & --- & 28.5 \\
   & 24.6 (82\%)  & 5.0 (93\%) & --- & --- & --- & 29.6 \\
   & 21.6 (72\%)  & 4.4 (89\%)  & 4.8 (95\%) & --- & --- & 30.8 \\
   & 16.4 (55\%) & 4.5 (91\%)   & 4.5 (90\%)  & 4.5 (89\%) &---& 29.9 \\
   & 11.8 (39\%)  & 4.2 (84\%)  & 4.4 (89\%) & 4.3 (87\%) & 10.9 (74\%)&35.6 \\
  \hline
\end{tabular}
\end{minipage}

\makeatletter\def\@captype{figure}\makeatother
\begin{minipage}{0.5\textwidth}
\centering
\includegraphics[scale=0.56]{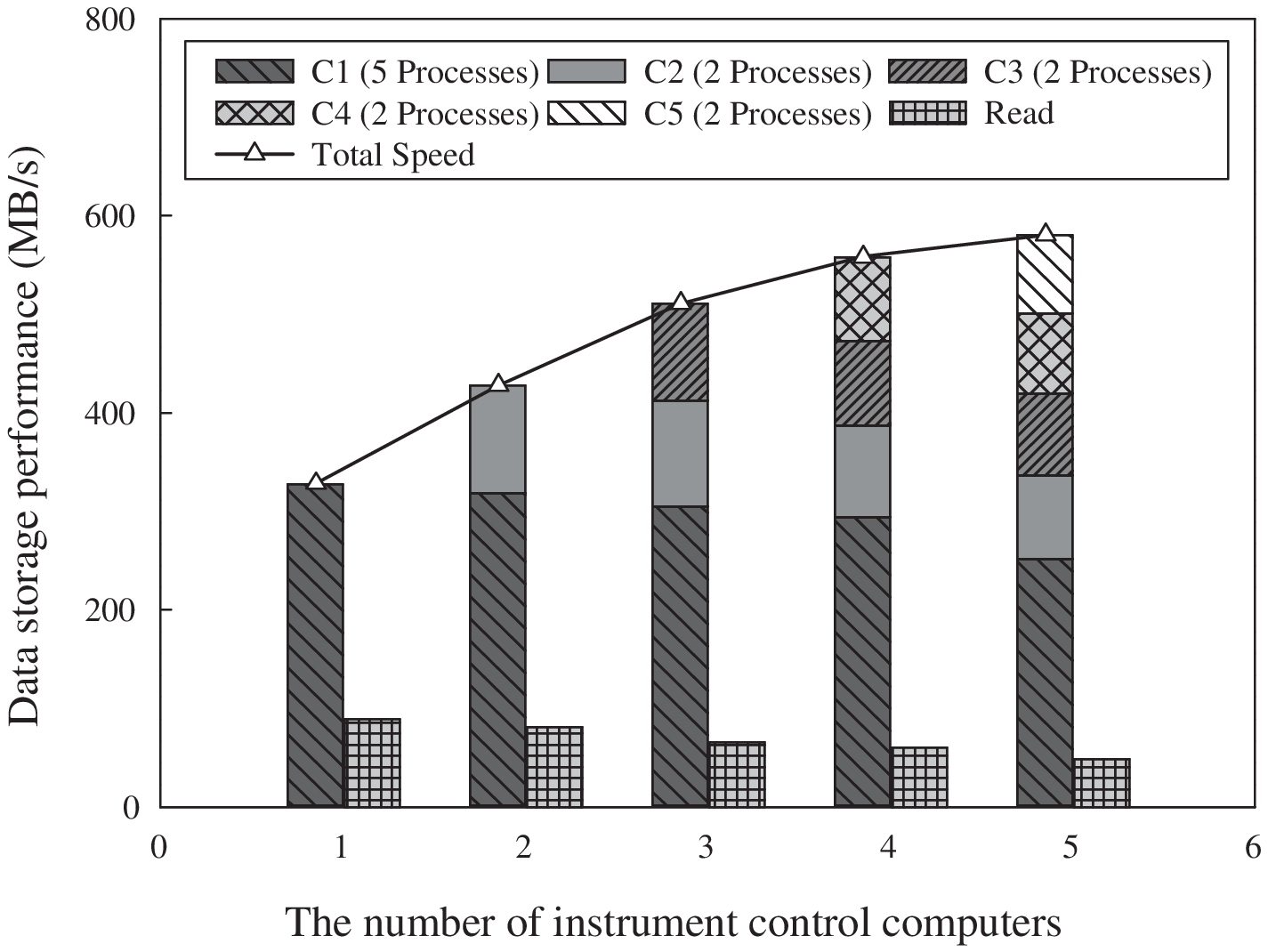}
 \caption{The write performance of hybrid storage strategy by using 64 MB FITS File and 5 processes for write. Meanwhile, a computer with 1 Gigabit Ethernet is used to read the data in parallel.\label{fig:hbperformance-a}}
\end{minipage}
\makeatletter\def\@captype{table}\makeatother
\begin{minipage}{0.45\textwidth}
 \caption{ Instruments data acquisition performance (unit: FPS) and the ratio of FPS (Real FPS / Maximum FPS)\label{tbl:hbperformance-b}}
\centering
\scriptsize
\tabcolsep 1pt
\renewcommand{\arraystretch}{1.3}
\vspace*{2mm}
\begin{tabular}{>{\centering}m{0.9cm}>{\centering}p{1.14cm}>{\centering}p{1.14cm}>{\centering}p{1.14cm}>{\centering}p{1.14cm}>{\centering}p{1.14cm}p{0.75cm}<{\centering}}
  \hline
  Cameras & Andor Neo (FPS) & PCO4000 (FPS) & PCO4000 (FPS) & PCO4000 (FPS) & PCO2000 (FPS)  & Total (FPS) \\
  \hline
  \multirow{4}{*}{Rates} & 29.8 (99\%) & --- & --- & --- & --- & 29.8 \\
   & 28.9 (96\%)  & 5.0 (100\%) & --- & --- & --- & 33.9 \\
   & 27.7 (92\%)  & 5.0 (100\%)  & 4.7 (94\%)& --- & --- & 37.4 \\
   & 26.7 (89\%) & 4.4 (89\%)   & 4.1 (82\%)  & 4.1 (81\%)&---& 39.3 \\
   & 22.9 (76\%)  & 4.0 (81\%)  & 4.0 (80\%) & 3.8 (77\%) & 9.9 (68\%)&44.6 \\
  \hline
  &  &   &  &  & & \\ 
\end{tabular}
\end{minipage}


\end{document}